\pdfoutput=1
\documentclass[aps,prl,twocolumn,superscriptaddress,showpacs]{revtex4}

\usepackage[ansinew]{inputenc}
\usepackage{graphicx}
\usepackage{fancyhdr}

\newcommand{\ieap}{Institut f\"ur Experimentelle und Angewandte Physik, Christian-Albrechts-Universit\"at zu Kiel, 24098 Kiel, Germany}

\newcommand{\tum}{Lehrstuhl f\"ur Theoretische Chemie, Technische Universit\"at M\"unchen, 85747 Garching, Germany}

\newcommand{\klpcn}{Key Laboratory for the Physics and Chemistry of Nanodevices, Department of Electronics, Peking University, Beijing 100871, P.~R.~China}

\newcommand{\ccme}{College of Chemistry and Molecular Engineering, Peking University, Beijing 100871, P.~R.~China}

\newcommand{\bir}{Beida Information Research (BIR), Tianjin 300457, P.~R.~China}

\newcommand{\hust}{School of Physics, Huazhong University of Science and Technology, 1037 Luoyu Road, Wuhan 430074, P.~R.~China}

\fancypagestyle{plain}{%
  \fancyhf{}%
  \fancyhead[r]{\footnotesize accepted for publication in Phys. Rev. Lett.}%
}

\begin{document}

\title {Spin Manipulation by Creation of Single-Molecule Radical Cations} 

\author{Sujoy Karan} \affiliation{\ieap}
\author{Na Li} \affiliation{\klpcn}
\author{Yajie Zhang} \affiliation{\ccme}
\author{Yang He} \affiliation{\klpcn}
\author{I-Po Hong} \affiliation{\klpcn}
\author{Huanjun Song}  \affiliation{\ccme}
\author{Jing-Tao L\"u} \affiliation{\hust}
\author{Yongfeng Wang} \email{yongfengwang@pku.edu.cn}
\affiliation{\ieap} \affiliation{\klpcn} \affiliation{\bir}
\author{Lianmao Peng} \affiliation{\klpcn}
\author{Kai Wu}  \affiliation{\ccme}
\author{Georg S. Michelitsch} \affiliation{\tum}
\author{Reinhard J. Maurer} \affiliation{\tum}
\author{Katharina Diller} \affiliation{\tum}
\author{Karsten Reuter}  \affiliation{\tum}
\author{Alexander Weismann} \affiliation{\ieap}
\author{Richard Berndt} \email{berndt@physik.uni-kiel.de} \affiliation{\ieap}



\begin{abstract}
All-\textit{trans}-retinoic acid (ReA), a closed-shell organic molecule comprising only C, H, and O atoms, is investigated on a Au(111)  substrate using scanning tunneling microscopy and spectroscopy. In dense arrays single ReA molecules are switched to a number of states, three of which carry a localized spin as evidenced by conductance spectroscopy in high magnetic fields. The spin of a single molecule may be reversibly switched on and off without affecting its neighbors. We suggest that ReA on Au is readily converted to a radical by the abstraction of an electron.
\end{abstract}

\maketitle
\thispagestyle{plain}

Tunable spins in adsorbed molecules are of interest for molecular spintronics applications~\cite{rocha_sanvito}. Transition metal complexes have often been used in this context ~\cite{bogani}, as their metal centers provide localized spins that may be protected by the surrounding organic framework. Chemical modifications to their ligand shells~\cite{spin1, spin2, spin3, spin4} were then found effective in manipulating spins of adsorbed molecules. 

Here we report scanning tunneling microscopy (STM) results for the biological  molecule retinoic acid (ReA, Fig.~1a) physisorbed to the inert Au(111) surface. Surprisingly, adsorbed ReA molecules may reversibly be converted among several states, some of which carry a localized spin as evidenced by conductance spectroscopy in high magnetic fields. A weakly bound adsorbate on a metal surface, such as ReA on Au(111), does \textit{a priori} not suggest the presence of such spin effects.  Previously, localized adsorbate spin and associated Kondo resonances have been found for adatoms~\cite{berndt_kondo,madhavan}, adsorbates comprising a metal center~\cite{hla, aitor, irspin6}, open-shell radical molecules~\cite{radical2} and systems in which significant charge-transfer between adsorbate and substrate occurs in the ground state~\cite{Torrente2008}. Such a ground-state charge transfer would not be expected for ReA. ReA also neither comprises a metal center, nor is it a radical. It is simply a closed-shell organic molecule. Nevertheless, single molecules in dense arrays may be switched to display a Kondo resonance without affecting their neighbors. We suggest that in these switched states the molecules are converted to a radical state by the removal of an electron from the endocyclic double bond of ReA\@. Indeed, stable radicals may preserve their spins upon adsorption on surfaces~\cite{radical1, radical3, radical4, radical2}. ReA and its derivatives exhibit great structural flexibility that living organisms use for signal transduction~\cite{bio_rea1}. We propose that this structural flexibility facilitates stabilization of the spin impurity upon switching.

Experiments were performed in ultra-high vacuum with three scanning tunneling microscopes operated at $\sim5$, $\sim4.3$, and $\sim0.5$~K\@.
Au(111) single-crystal surfaces and etched tungsten tips were prepared by Ar ion bombardment and annealing cycles. ReA was sublimated onto Au(111) at ambient temperature in ultrahigh vacuum and imaged at approximately $\sim0.5 - 5$~K\@. Images are displayed as illuminated three-dimensional surfaces. For detection of the differential conductance, $dI /dV,$ we applied standard lock-in techniques. Calculations were carried out using dispersion-corrected density-functional tight-binding (DFTB) as implemented in Hotbit~\cite{Koskinen2009} and dispersion-corrected GGA-PBE \cite{pbe} density-functional theory (DFT) as implemented in FHI-AIMS~\cite{Blum2009}, both in combination with the ASE package~\cite{Bahn2002}. Dispersion interactions were supplied by the vdW\textsuperscript{surf}-method~\cite{Ruiz2012}. Charge resonant excited states were calculated using the le$\Delta$SCF approach~\cite{Maurer2013} as implemented in a local version of CASTEP 6.0.1~\cite{castep}. More details are available in the Supplemental Material \cite{suppm}.

\begin{figure} 
\includegraphics[width=0.98\linewidth]{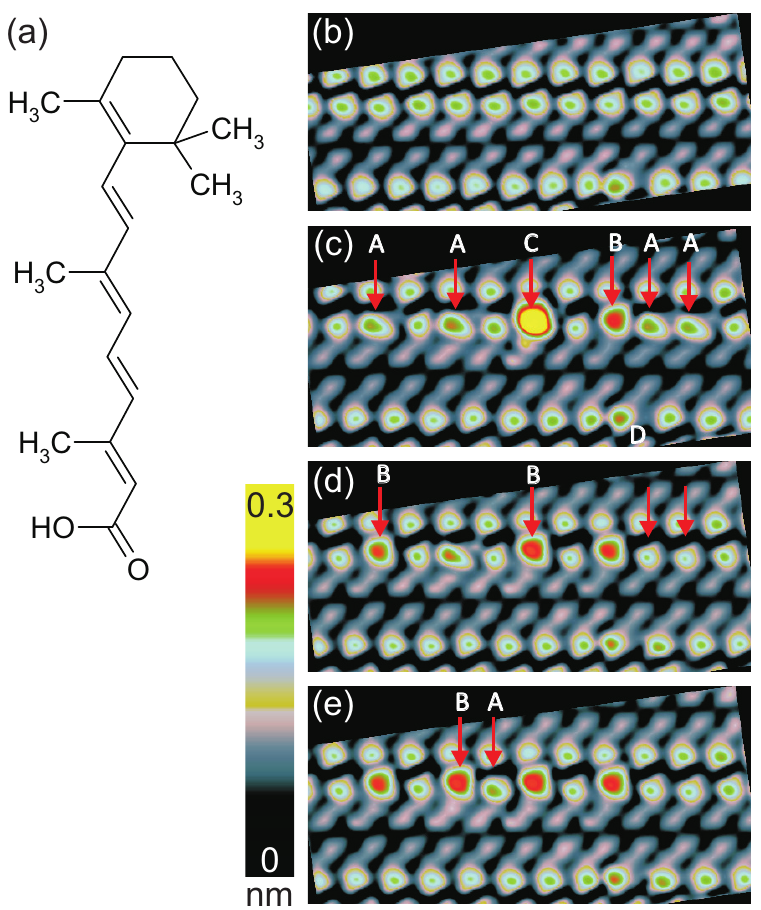} 
\caption{ 
(a) Schematic of all-\textit{trans}-retinoic acid (ReA).
(b) STM image ($9.7 \times 4.3$~nm$^2$) of striped pattern comprising ReA dimers on Au(111), acquired at $\sim 5$~K\@. The molecules are in their pristine form as deposited. (c) STM image of the same area after manipulating selected molecules at a sample voltage $V=-2.5$~V\@.
Only the manipulated molecules are indicated with red arrows. The different states obtained are labeled A--D\@. (d, e) Further manipulation steps. Only those molecules that have been modified compared to the previous image are labeled. All images were acquired with $V = -5$~mV at a constant current $I = 100$~pA\@. The color scale covers a range of 0.3~nm. The image contrasts were essentially unchanged within the range $-2$~V $<V<$ 2~V\@.}
\end{figure}

\begin{figure}
\includegraphics[width=0.7\linewidth]{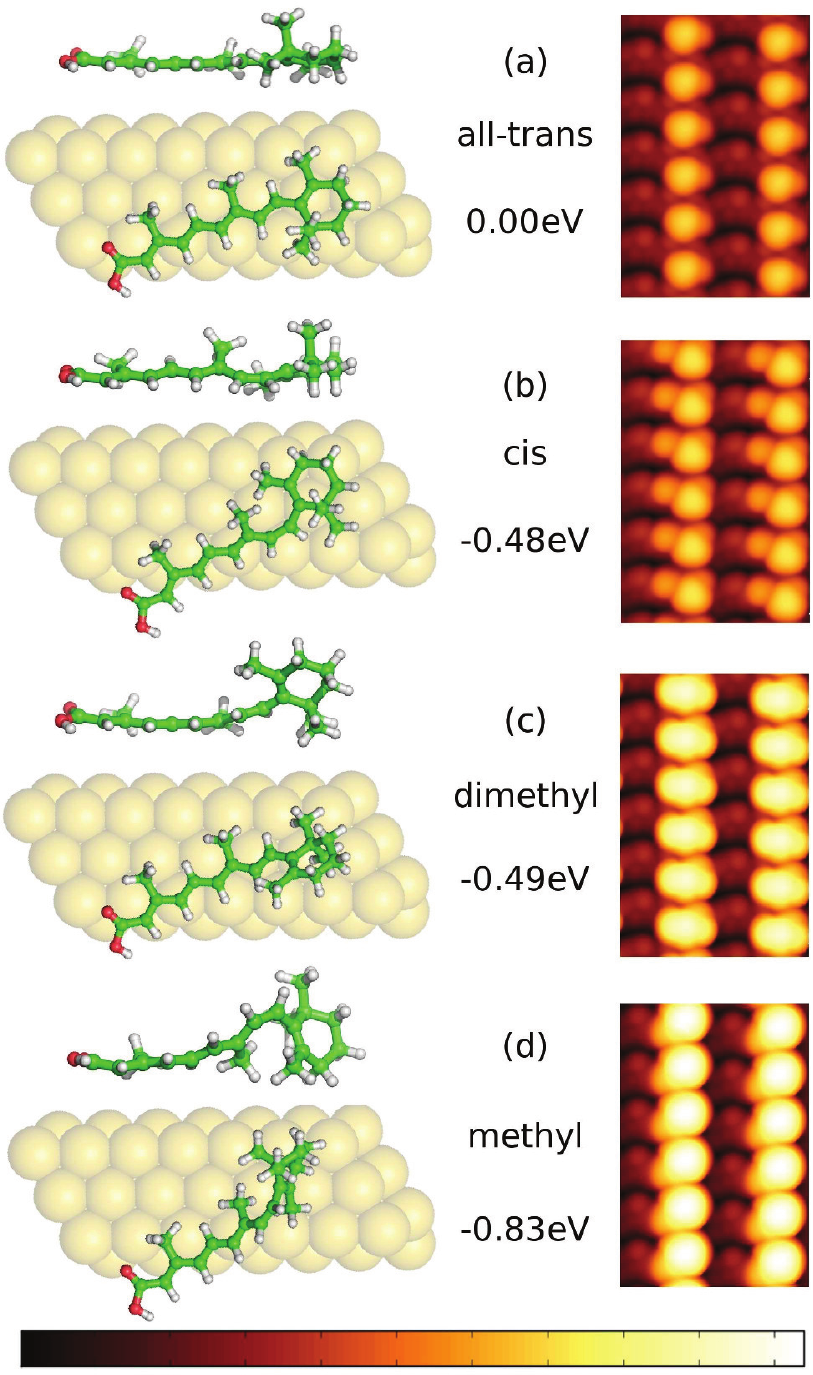}
\caption{
Left: Top and side views of the geometries of the DFT+vdW$^{\mathrm{surf}}$ optimized conformers, denoted all-\emph{trans}, \emph{cis}, dimethyl, and methyl. Right: Corresponding Tersoff-Hamann simulated STM topographs ($-2.5$~V) and relative decrease in adsorption energy with respect to the most stable all-\emph{trans} configuration. The color scale covers apparent heights from 0.5 to 10.4~nm (dark to bright). The conformational characteristics (\textit{e.~g.}, the uplifted head group in dimethyl) are reflected in the topographic images.}
\label{fig:stm} 
\end{figure}

Close to monolayer coverage, ReA forms a dense pattern that covers large areas of the Au(111) surface (Fig.~S1 of the Supplemental Material \cite{suppm})~\cite{sk_rea}. Figure~1b shows a detailed image revealing that ReA molecules arrange themselves in parallel dimers. The highest protrusions correspond to the bulky 1,3,3-trimethylcyclohexene groups (henceforth denoted as head groups) and appear $\sim 60$~pm higher than the rest of the molecules (3,7-dimethylnona-2,4,6,8-tetranoic acid, henceforth denoted as polyene). Slight variations in contrast among molecules within the array are likely due to variations of the adsorption site on the atomic lattice of Au(111). Such an incommensurability is not surprising because the selective and directionally defined H-bonds mediated by the carboxylic groups may prevail over interfacial van der Waals interactions in attaining the eventual structure. No preferred orientation of the molecular patterns on the substrate was observed, which is an indication of a limited corrugation of the substrate potential. This finding is furthermore supported by DFT calculations of the adsorbed molecule (cf.\ Section~2.3 of the Supplemental Material \cite{suppm}), where no strong energetic preference for a specific adsorption site was found.

ReA molecules in arrays can be switched selectively among different states by placing the tip over the neck of the bulky heads and lowering the sample voltage to $-2.5$~V with current-feedback disabled. The process was effected by monitoring the current through the molecular junction, which switched back and forth among well-defined levels, and increasing the voltage (to values closer to 0 V) when the current corresponding to a desired state was detected. Figure~1c shows an intermediate upshot. Three distinct shapes of the manipulated states, denoted A, B, and C (Fig.~1c), surrounded by pristine molecules, are discernible. While the head group in state A appears as an elliptical protrusion, a rounded Reuleaux triangular shape is viewed in state B\@. State C displays an extended protrusion along with a kink in the polyene chain at the back of the neck of the bulky head. Switching was found to lead to some further states, the most common one of which is labeled D in Fig.~1c.

The manipulated states were found to be stable over periods as long as days at the temperature of the experiments (5~K). Spontaneous conversion was never observed so long as nondestructive currents and voltages (usually $|{V}|\le 2$~V) were used for imaging. The degree of selectivity in attaining a particular state is further illustrated in Figs.~1d and e. First, different states were prepared in Fig.~1d (keeping only two manipulated molecules of Fig.~1c unchanged), next most manipulated molecules have been converted to state B in Fig.~1e. The differences in apparent heights among different states is clearly visible in the pseudo three-dimensional images and line profiles taken along the bulky heads of manipulated molecules (Supplemental Fig.~S2). The head groups of states A, B, and C appear higher ($\sim 20$, $\sim 60$, and $\sim 140~$pm, respectively) relative to pristine molecules \cite{NOTE}. This substantial increase in apparent heights indicates either a strong conformational change of the molecule or a redistribution of the local density of states or a combination thereof. In contrast, the tails of the polyene chains remain essentially unchanged.

To identify possible geometric arrangements we employed an extensive DFTB/DFT+vdW$^{\mathrm{surf}}$~\cite{Ruiz2012} screening approach (see Sect.~2.1 of the Supplemental Material \cite{suppm} for details). We found four distinct groups of stable adsorption geometries, for which we also simulated corresponding STM images (Fig.~\ref{fig:stm}). Taking into account relative adsorption energies and comparing simulated and measured STM data, we conclude that in its unswitched ground state ReA adsorbs in an all-\textit{trans} configuration. Consistent with the geometrical interpretation of the experimentally observed switched states, the other three conformational groups identified (designated as \textit{cis}, methyl, and dimethyl in Fig.~\ref{fig:stm}) all exhibit an uplifted and tilted head group, \textit{i.~e.}, the carboxy group anchors the molecule to the substrate, whereas the remaining molecule (especially the ring system) is lifted away from the surface facet and stabilized due to a tripodal arrangement of methyl-substituent groups and the conjugated chain. Differing in the orientation of the ring system, the three conformers yield simulated Tersoff-Hamann STM images with either elliptical or round protrusions, reminiscent of the different appearances of the switched states in the experiment. We correspondingly conclude that the experimentally observed topographic changes are not only rooted in a change in electronic structure, but also arise from profound changes in geometry caused by the electronic stimuli.

\begin{figure} 
\includegraphics[width=0.89\linewidth]{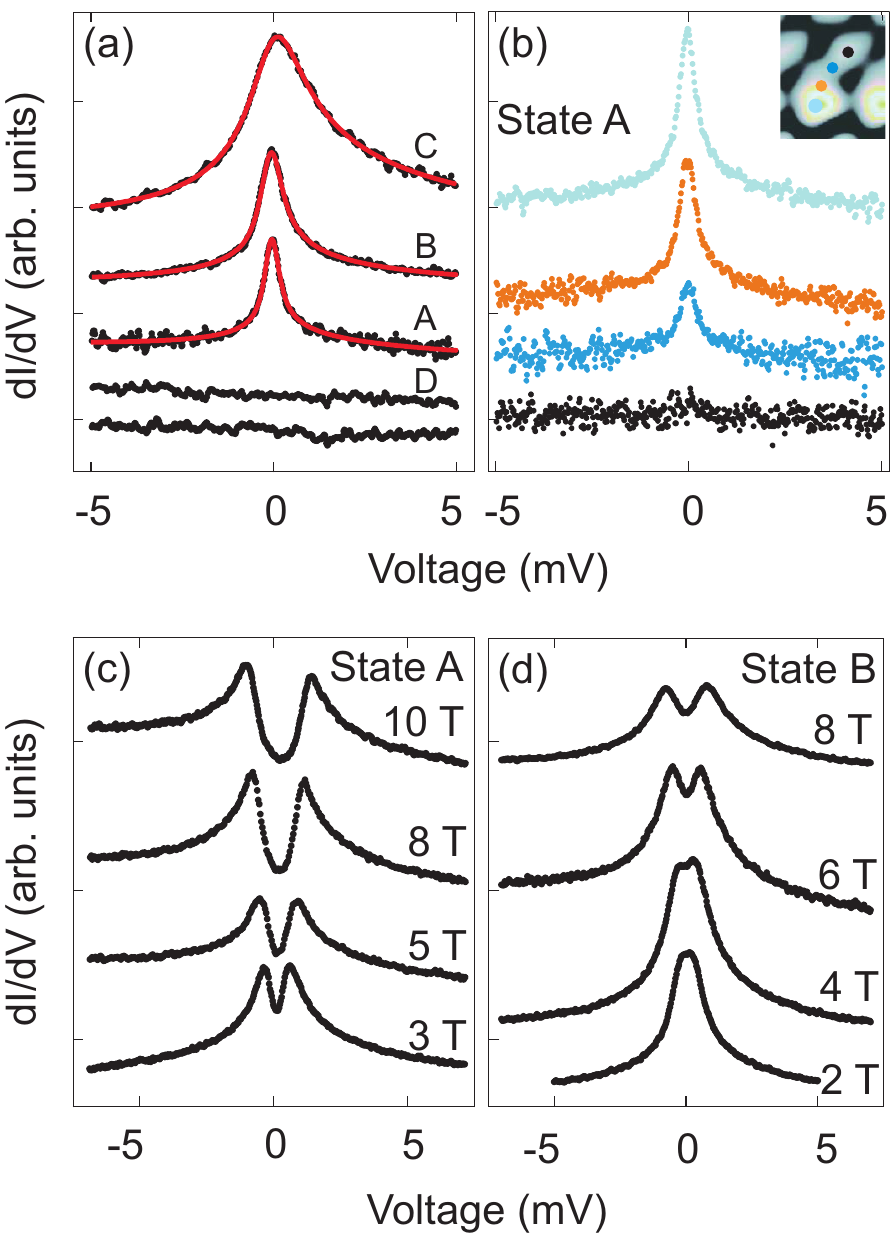} 
\caption{ 
(a)	$dI/dV$ spectra recorded over the center of the bulky heads of pristine ReA (bottom spectrum) and different manipulated states (labeled A--D). Before opening the current feedback for spectroscopy, the STM was operated at $V = -10$~mV, $I =	100$~pA, and $T=0.5$~K. Vertical offsets are used for clarity. Red curves represent fits of Frota functions to the spectra (see Section~1.6 of the Supplemental Material \cite{suppm} for details). (b) Spatial evolution of $dI/dV$ spectra along the polyene chain of state A\@. Colored dots in the inset indicate the positions at which the spectra were recorded. For clarity, the spectra are shifted vertically. Similar results were obtained for other manipulated states.
(c, d) $dI/dV$ spectra of states A and B in magnetic fields at $T=0.5$~K\@. The field strengths are indicated in the figure. Splitting of the Kondo resonance of state A is observed at all field strengths. State B displays some splitting at approximately 4 T\@.
\label{spex} }
\end{figure}

The electronic properties of all five states of the molecule mentioned above are embodied in the differential conductance ($dI/dV$) spectra shown in Fig.~\ref{spex}a. The featureless spectrum (bottom curve) was recorded from a pristine molecule. Other spectra taken over the center of the bulky heads of the manipulated states surprisingly show a zero-bias resonance (except for state D), similar to the Kondo fingerprint reported from atoms and molecules on surfaces~\cite{berndt_kondo, madhavan, irspin1, hla, aitor, irspin6, irspin9, dilullo}.
As the tip is moved sideways along the polyene chain, the resonance vanishes (Fig.~\ref{spex}b).

To determine the origin of the zero bias anomaly, the evolution of $dI/dV$ spectra was examined in the presence of a variable magnetic field $B$. Figure~\ref{spex}c shows a series of spectra from the head group in state A recorded at various field strengths. As expected for a Kondo resonance, the zero-bias peak splits as $B$ is increased. Similar results were obtained for state B (Fig.~\ref{spex}d). In state C (Supplemental Fig.~S3) no splitting was achieved at the maximum $B$ available at the time of the experiment. However, the involvement of the molecular spin is evident from the  broadening at $B=6$~T\@. These data exclude the possibility that the feature at the Fermi level $E_F$ is due to the excitation of low energy vibrations~\cite{Schneider, Hofer}. Information on the Kondo temperatures and the fitting procedures are provided in Sect.~1.6 of the Supplemental Material \cite{suppm}. We note that multistability was previously reported from tetracyanoethylene on Cu(111) with one state displaying several low-energy excitations~\cite{gupta}. An excitation around $E_F$ has been attributed to a Kondo resonance, but no field dependence was reported.

The Kondo resonance indicates the presence of a majority spin located at the head group of the molecule, which is only observed after the manipulation of the pristine all-\textit{trans} configuration. At neither semi-local (PBE \cite{pbe}) nor hybrid (PBE0 \cite{pbe0}) DFT level of theory do we find any significant localization of excess spin for the different ReA isomers in the gas phase. The localized spin giving rise to a Kondo feature must therefore be a direct result of the adsorbate-substrate interaction. In light of the experimental observations we therefore suggest that the STM manipulation converts the molecular adsorbate to a charged radical. As switching to states that exhibit a Kondo resonance was only observed at negative sample voltages, this is specifically the formation of a positively charged radical (ReA*$^+$). This interpretation is consistent with the findings for solvated ReA, where addition of an electron leads to dissociation via deprotonation \cite{deprotonation}, whereas electron removal leads to an equally stable radical cation under acidic and alkaline conditions \cite{Rea_radical}.

The change in the molecular structure observed in the STM images is likely important in stabilizing the charged molecule, similar to the case of charged metal atoms on insulating films \cite{Repp_1,Repp_2,charge_state_3}. In a similar way, the geometrical distortion of ReA upon electron withdrawal is likely to dissipate excess energy into the substrate, thereby effectively trapping the metastable ionic state.
The fact that the strongest Kondo feature is observed at the cyclohexene group suggests that the charge is localized close to the alkene carbon atoms, far away from the carboxy group that anchors the molecule to the metal substrate. Stable charged states were indeed previously reported from a monolayer of charge-transfer complexes on Au(111) \cite{Torrente2008}. Furthermore, it has been shown that carotenoid cations in solution are stable over 6~h at 250~K \cite{Lutnaes2004}. The formation of a long-lived cationic state on the surface thus appears to be a viable proposition.

We experimentally also investigated ReA on Ag(111), Cu(111), and Pb(111) surfaces (Supplemental Material \cite{suppm}, Sect.~1.5). On none of these substrates ReA could be switched to display a Kondo effect. On the other hand, for 13-\textit{cis}-retinoic acid, an isomer of ReA, we succeeded in preparing spin-carrying states similar to ReA on Au(111) (Supplemental Material \cite{suppm}, Sect.~1.4). Apparently, the weak coupling of ReA to the inert Au(111) substrate is required to enable structural flexibility and radical formation.

Semi-local DFT-PBE calculations of the adsorbed ReA conformers did not exhibit any degree of spin localization on the molecule. This is likely a consequence of the electron delocalization problem at this level of theory \cite{Stroppa2008}. Adequately capturing both the molecular frontier orbitals and the metal band structure presumably requires screened-exchange DFT functionals, which are presently untractable for system sizes like ReA at Au(111). To nevertheless further explore the charged states, we generated excited states of the four optimized conformers on the surface by selectively removing (adding) an electron from the highest occupied molecular orbital (into the lowest unoccupied molecular orbital, LUMO). Rather independent of the actual molecular conformation, the energy to remove an electron from the adsorbed molecule is obtained as $\sim$ 0.8~eV, and therewith approximately half the energy needed to add an electron to the LUMO (see Supplemental Table S2). This suggests that formation of a cationic state is indeed feasible at the bias voltages used in the experiments to induce switching ($-2.5$~V). More importantly, the addition energy is close to its analogue in the gas-phase, the electron affinity. In contrast, the electrostatic interaction between the molecular charge and its image charge strongly reduces the electron removal energy of the adsorbed molecule compared to the ionization potential in gas phase ($\sim$ 6.7~eV). This further supports that the formation of a cationic radical is specifically facilitated at the surface.

In summary, retinoic acid molecules on gold may be converted to states that carry a localized spin. This suggests that a corresponding spin localization is not restricted to molecules containing metal centers, but instead extends to biostable and initially non-magnetic molecules.
We propose that a radical cation is created by the removal of an electron from the endocyclic double bond. A possible ramification of this work is to use arrays of ReA and related molecules to create arbitrary patterns of spins. While the direct interaction of these spins may be small the underlying metal surface may mediate long-range interactions.

We thank Roberto Robles and Nicol\'as Lorente for discussions. SK, YW, AW, and RB acknowledge financial support from the DFG via the SFB 677.
NL, YZ, YH, IH, HS, JL, YW, LP, and KW acknowledge the National Natural Science Foundation of China (21522301, 21373020, 21403008, 61321001, 21433011, 21333001), Ministry of Science and Technology (2014CB239302, 2013CB933400), and the Research Fund for the Doctoral Program of Higher Education (20130001110029). RJM and KR gratefully acknowledge financial support from the DFG\@. Computer resources for this project have been provided by the Gauss Centre for Supercomputing/Leibniz Supercomputing Centre, grant pr94sa.

\end{document}